\documentstyle[prl,aps]{revtex}
\begin{document}
\draft

\title{General teleportation channel, singlet fraction and quasi-distillation}

\author{Micha\l{} Horodecki \cite{poczta1}}

\address{Institute of Theoretical Physics and Astrophysics\\
University of Gda\'nsk, 80--952 Gda\'nsk, Poland}

\author{Pawe\l{} Horodecki\cite{poczta2}}

\address{Faculty of Applied Physics and Mathematics\\
Technical University of Gda\'nsk, 80--952 Gda\'nsk, Poland}

\author{Ryszard Horodecki \cite{poczta3}}

\address{Institute of Theoretical Physics and Astrophysics\\
University of Gda\'nsk, 80--952 Gda\'nsk, Poland}

\maketitle

\begin{abstract}
We prove a theorem on direct relation between the optimal fidelity $f_{max}$
of teleportation and the maximal singlet fraction $F_{max}$ attainable by means
of trace-preserving LQCC action (local quantum and classical communication).
For a given bipartite state acting on $C^d\otimes C^d$ we have
$f_{max}= {F_{max}d+1\over d+1}$. We assume completely general teleportation
scheme (trace preserving LQCC action over the pair and the third particle in
unknown state). The proof involves the isomorphism between quantum channels and 
a class of bipartite states.  We also exploit the technique of $U\otimes U^*$ 
twirling states (random application of unitary transformation of the above 
form) and the introduced analogous twirling of channels. We illustrate 
the power of the theorem by showing that {\it any} bound entangled state
does not provide better fidelity of teleportation than for the purely
classical channel. Subsequently, we apply our tools to the problem of
the so-called conclusive teleportation, then reduced to the question of optimal
conclusive increasing of singlet fraction. We provide an example of
state for which Alice and Bob have no chance to obtain perfect singlet
by LQCC action, but still singlet fraction arbitrarily close to
unity can be obtained with nonzero probability. We show that a slight
modification of the state has a threshold for singlet fraction which cannot be
exceeded anymore.
\end{abstract}

\section{Introduction}
Consider the following problem.
Alice and Bob are far from each other, and they share one
pair of spin-$s$ particles in entangled quantum state $\varrho$.
All the manipulations Alice and Bob are alowwed to perform are
local quantum operations and classical
communication (called LQCC or bi-local operations). It means that
that, in particular, they cannot
exchange quantum bits or
establish quantum interaction between their labs.
Now, suppose that Alice wants to teleport an unknown state of some other
spin-$s$ particle, but only if she is sure that the fidelity $f$ of the
transfer is better than  some given threshold $f_{min}$ ($f_{min}<1$).
Which states $\varrho$ give Alice nonzero chance
that after some LQCC operations she can teleport being sure
that the her requirement is satisfied?
The answer to this question will be one of the
result of the presented paper.


As one knows,  quantum teleportation \cite{Bennett_tel}
allows to transfer the quantum information through quantum entangled
states as quantum channels (supported by classical channels) with the
fidelity better than by means of classical channel itself \cite{Pop}.
For example, two spin-$s$ particles in the maximally entangled state
\begin{equation}
P_{+}=|\Psi_{+} \rangle \langle \Psi_{+} |, \ \
|\Psi_{+} \rangle=\frac{1}{\sqrt{d}}\sum_{i=0}^{d} |i\rangle |i\rangle, \ \ d={2s+1}
\label{max}
\end{equation}
(we shall call it singlet state, despite it is in fact local transformation of
true singlet)
shared by a sender - Alice - and a receiver - Bob
allows to transmit faithfully an unknown spin-$s$ state, with additional use
of classical bits  describing one of $(2s+1)^{2}$ elementary messages.
If the state shared by Alice and Bob
is pure but not maximally entangled than
one can perform {\it conclusive} teleportation \cite{Tal}.
The main idea of the latter is that, given a
particle in {\it non-maximally} entangled state $\Psi$ providing
small transmission fidelity $f$, Alice and Bob can transform
the state by some deliberate LQCC operations.
As a result, with some probability the final state provides much greater
transmission fidelity (usually the perfect one).
In the proposed protocol \cite{Tal} the pairs of particles were treated
{\it non-collectively} i.e. each pair was processed separately.
The concept of {\it collective} operations that involve interactions among
different pairs, has been implemented for pure states in the protocol
of {\it concentration
of entanglement} by LQCC operations \cite{conc} (see \cite{Rohrlich}
for interesting consequences for entanglement measures
and \cite{Hoi-Kwong} for analysis of probability distributions).
It this approach some number of non-maximally entangled states
is converted via LQCC operations into the less number
of maximally entangled ones which can be used for instance
for faithful teleportation process.

In realistic conditions, instead of pure entangled
states Alice and Bob usually share mixed state that contains noisy
entanglement.
The latter case is more complex and it has longer history.
Popescu first has pointed out \cite{Pop}
that mixed entangled states can allow for teleportation with
significantly better fidelity than the one achieved by using
only classical bits.
He also showed \cite{hid}  that some non-collective
LQCC operations can transform the $d \times d$ ($d\geq 5$)
Werner entangled mixed states satisfying local hidden
variable model \cite{Werner} into the two spin-${1 \over 2}$ states
violating Bell inequalities. Subsequently, similar effect by means of
local filtering  for mixed two spin-${1 \over 2}$ states has been found
\cite{Gisin} (see \cite{conc} for pure state case).
At the same time the important idea of
{\it distillation} (or {\it purification}) of noisy entanglement
has been worked out \cite{Bennett_pur,Deutsch}.
Here the aim is to convert some number of {\it mixed}
inseparable states into less number of states close to
maximally entangled pure one. The distillation protocols are
usually accomplished by operating on collections of pairs rather than on
single pairs. However, the single pair operations introduced
in Refs. \cite{Gisin,conc} have been shown
to play an important role in the distillation protocol
capable to distill {\it all} entangled two-qubit states \cite{pur}.
On the other hand, single pair operations are
much  simpler to perform experimentally \cite{Gisin,zuk}.
First general results concerning the limits for those operations have
been provided by Linden, Massar and Popescu \cite{Linden} and
by Kent \cite{Kent}.

In this paper we would like to consider the question concerning the conclusive
teleportation we asked at the beginning. We first reduce the question
to the problem of single pair distillation. To this end we provide a
number of tools which can be useful also in more general context.

In sec. \ref{isomorphism} we consider the problem of equivalence between bipartite states and
quantum channels. The connections between states and channels were considered
in Refs. \cite{huge,Schumacher}.
It is clear that if we have channel, then we can produce a
bipartite state sending a half of singlet down the channel \cite{huge}.
However, given a
state, it is not clear, whether there is a channel, which can produce it in
the above way. A way of ascribing channel to a given state is to perform
teleportation via the state (creating the teleportation channel)
One can now ask what channels can be produced by means of
teleportation via a given mixed state. Another question is the following
\cite{huge}:
suppose that a mixed state was produced from a channel, by sending half of
singlet. Can we recover the channel by means of some (probably very
sophisticated) teleportation scheme applied to the state? This is the question
of reversibility of the operation of producing states from channels.
In sec. \ref{isomorphism} we prove that the operation of sending
half of singlet down the
channel produces isomorphism between channels states having the reduced
density matrix of one of  subsystems maximally chaotic (proportional to
identity).

The section \ref{fidelity-fraction} is devoted to the problem of optimal
fidelity of teleportation. We first consider
families of states and channels connected
 via the above isomorphism. The states are singlets with admixture of
completely mixed state \cite{xor} (generalization of two-qubit Werner states
\cite{Werner,Pop}) which are the only states invariant under the $U\otimes
U^*$ twirling introduced in Ref. \cite{xor}.
The channels are the generalized depolarizing ones.  We show that the families
are not only connected via the isomorphism, but are also physically
equivalent, i.e. the teleportation via state reproduces the channel.
This generalizes similar observation for two-qubit Werner states \cite{huge}.
We also introduce an operation on channels which is equivalent to twirling
of the corresponding states. We show that average fidelity
and the entanglement fidelity are invariants of the twirling operation.

These are the main tools which allow us to prove a strict connection between
the optimal fidelity $f_{max}$ of teleportation via a given state and the maximal
singlet fraction $F_{max}$ attainable by means of trace preserving LQCC operations
on single pair. Namely we prove that there holds the equality
$f_{max}={F_{max}d+1\over d+1}$. We emphasise here, that we consider the most
general teleportation scheme which is possible. Then the problem of optimal
teleportation fidelity is reduced to the much less complicated (but still
nontrivial) task of optimal increasing singlet fraction by means of
trace-preserving LQCC operations. We illustrate the power of the result
in sec. \ref{BE}
applying it to the problem of optimal teleportation fidelity
via bound entangled states (the entangled ones which cannot be distilled
\cite{bound}). This problem has been quite recently risen by Linden and
Popescu \cite{Lind_ost} who showed that some of bound entangled states do not
provide better fidelity by teleportation via purely classical channel
(see \cite{Pop} in this context). Here we prove that it is true for any bound
entangled states and for the most general teleportation schemes.

In subsequent sections we apply the results to the problem of conclusive
teleportation which is now reduced to the problem of conclusive increasing
singlet fraction. We consider two concepts: (i) non-collective distillation,
where Alice and Bob have a chance to obtain a perfect singlet and (ii)
non-collective quasi-distillation (in short quasi-distillation), where
the perfect singlet cannot be obtained, but there is nonzero chance to obtain
arbitrarily high singlet fraction. From the results concerning
non-collective distillation \cite{Linden,Kent} we know that for
a broad class of mixtures it is impossible even to increase the singlet
fraction. As shown in Ref. \cite{Kent}  the non-collective distillation is
impossible for mixed states of full rank.

Here we address, in particular,  the following question: does there
exist a state which is not non-collectively distillable, but still is
quasi-distillable? To answer this question we determine (in sec.
\ref{distillation})
the class of the
states which can be distilled (we generalize the consideration taking into
account the case of singlet of less dimension than the dimension of the
system: this corresponds to fidelity teleportation with the average calculated
over inputs restricted to less Hilbert space). In sec. \ref{quasi-distillation}
we provide example of
state which does not belong to this class, but can be quasi-distilled.
Then we slightly modify the state so that the new one cannot be
quasi-distilled having a threshold  for singlet fraction which cannot be
exceeded. Thus for sufficiently high required fidelity of teleportation, Alice
and Bob have no chance to obtain the fidelity for this state.

\section{States and channels}
\label{isomorphism}
In this section we prove formally that the set
of channels $\Lambda$ on the set of $d$-dimensional
states is isomorphic to the set of density matrices $\varrho$ acting on
the Hilbert space
${\cal H}={\cal H}_1\otimes {\cal H}_2=C^d\otimes
C^d$ satisfying ${\rm Tr}_{{\cal H}_2} \varrho={I\over d}$ (the partial trace
over the second system gives maximally mixed state).
By channel we mean here completely positive,
trace-preserving map \cite{CP}. Given a channel $\Lambda$ one can ascribe to it a
state $\varrho_\Lambda$, sending half of singlet state down the channel
\begin{equation}
\varrho_{\Lambda}=(I\otimes\Lambda)P_+.
\label{map}
\end{equation}
Such a state must have the first reduction the same as the state $P_+$
(i.e. ${I\over d}$) since, due to impossibility of action at a
distance the local reduced density matrix of the remote half of $P_+$ cannot be
changed by any local action performed on the other half (this is a
``physical'' version of the proof of this
well known fact \cite{Jamiolkowski}).

Now consider a given state $\varrho$, with first reduction equal to
${1\over d}I$. Following   Ref. \cite{xor}, consider the
spectral decomposition of the state
\begin{equation}
\varrho=\sum_{i=1}^{d^2}p_k |\psi_k\rangle\langle \psi_k|.
\end{equation}
Let, e.g., $\psi_1=\sum_{i,j=1}^d c_{ij} |i\rangle\otimes |j\rangle$. Then it
can be represented as
\begin{equation}
\psi_1=I\otimes V_1\psi_+
\end{equation}
where $\langle i|V_1|j\rangle=\sqrt{d}c_{ij}$.
Defining analogously $V_k$ for $k=1,...,d^2$ we obtain
\begin{equation}
\varrho=\sum_{k=1}^{d^2}p_k I\otimes V_k P_+ I\otimes V_k^{\dagger}=
(I\otimes \Lambda)P_+
\label{lipa}
\end{equation}
where $\Lambda(\sigma)=\sum_kp_kV_k\sigma V_k^\dagger$. Of course
the map $\Lambda$ defined in this way
is completely positive, since it is of the common Stinespring form
\cite{Stinespring}. It is also trace-preserving. To show it we only need to
check whether $A\equiv\sum_k p_k V_k^\dagger V_k=I$ \cite{Jozsa}.
Since the first reduction of our state $\varrho$ is ${1\over d}I$, we have
for any operator $B$
\begin{equation}
{\rm Tr} B=d{\rm Tr}(\varrho B\otimes I)=
d \sum_kp_k{\rm Tr} (P_+ B\otimes V_k^\dagger V_k)=d {\rm Tr} (P_+B\otimes A)
\end{equation}
Now using the property that $C\otimes IP_+=I\otimes C^TP_+$  for any operator
$C$ and the fact that the reduction of the singlet is ${1\over d}I$ we obtain
\begin{equation}
{\rm Tr} B={\rm Tr}B^TA=TrA^{T}B
\end{equation}
for any $B$.
This implies that $A^T=I$ hence of course also $A=I$.

Finally one should know that the channel $\Lambda$ is determined uniquely.
Suppose, that there are two maps $\Lambda$ and $\Lambda'$ which produce
the same state $\varrho$ so that we have
\begin{equation}
(I\otimes(\Lambda-\Lambda'))P_+=0
\label{zero}
\end{equation}
Denote the difference $\Lambda-\Lambda'$ by $\Gamma$. We will now show that
$\Gamma$ must be equal to $0$. Indeed, consider the operator basis constituted
by the following operators $P_{ij}=|i\rangle\langle j|$. In that basis we have
$P_+={1\over d} \sum_{ij}P_{ij}\otimes P_{ij}$. Substituting it into
formula (\ref{zero}) we obtain
\begin{equation}
\sum_{ij}P_{ij}\otimes \Gamma(P_{ij})= \sum_{ijkl}
\gamma_{ijkl} P_{ij}\otimes P_{kl}=0
\end{equation}
where $\gamma_{ijkl}$ are matrix elements of $\Gamma$ in the basis $P_{ij}$.
Since the  operators $P_{ij}\otimes P_{kl}$ also  constitute basis
then it follows that all $\gamma_{ijkl}$ must vanish so that $\Gamma$
must be $0$ operator. Then, we have shown that if two maps  give raise
to the same state $\varrho$ then they must be equal, so that the formula
(\ref{map}) determines $\Lambda$ uniquely.
Thus it constitutes an affine  {\it isomorphism} between channels and
states of maximally mixed reduced density matrix on one of the subsystems.

Having proved the isomorphism, given a state $\varrho$ we will denote by
$\Lambda_{\varrho}$ the unique channel satisfying (\ref{map}). Conversely,
given a channel $\Lambda$ we ascribe to it a state $\varrho_{\Lambda}$
also by means of the formula (\ref{map}).
Note that so far this isomorphism has been established here
only between channels i.e. completely
positive trace-preserving maps and quantum
states with one of subsystems being completely mixed.
One can easily see that  the isomorphism can be extended to
all states if one abandon the condition of preserving the trace.
Then we have the one-to-one correspondence between
the set of all states and the set of all completely positive  maps.
But we would like to stress here that
we refer to a completely positive
map as to a channel only if it is trace-preserving.

Let us now discuss the physical sense of the considered useful mathematical
equivalence. If Alice and Bob are connected via a channel $\Lambda$
then they can create state $\varrho_\Lambda$ by sending half of singlet
down the channel. However, if they initially share the state
$\varrho_\Lambda$, then  can they say they dispose a channel
$\Lambda$?
As one knows, applying teleportation protocol, they, in fact, obtain some
quantum channel. It remains an open question whether there exists some
teleportation procedure which {\it reproduce} the channel $\Lambda$ \cite{huge}.
It is
highly probable that, in general, sending a half of singlet down the channel
causes some {\it irreversible} lost of the capacity. Thus, the mathematical
equivalence would not imply the physical one in general. One knows that
in some cases there is also a physical equivalence. Namely, for the two-qubit
Werner state the corresponding channel (quantum depolarizing channel) can be
retrieved by applying the standard teleportation protocol \cite{huge}.
In the next section
we will show that the same reversibility holds for generalized depolarizing
channel (associated with the family of the $U\otimes U^*$ invariant states).

At the end of this section we define two parameters describing channels and
states. We will use the same notation for states and for channels, but
the parameters will of course have different interpretation. We will denote
them by $f$ and $F$. The first one is defined for channels in the
following way
\begin{equation}
f(\Lambda)=\int{\rm d}\,\phi \langle\phi
|\Lambda(|\phi\rangle\langle\phi|)|\phi\rangle,
\label{fid2}
\end{equation}
where integral is performed with respect of the uniform distribution
$d\phi$ over all input pure states.
It has the following interpretation: it is the probability
that the output state
$\Lambda(|\phi\rangle\langle\phi|)$ passes the test of being the input state
$\phi$, averaged over all input states. We will call it  fidelity of
the channel.

For the states, the parameter $f(\varrho)$ will denote the fidelity of the
channel constituted by
the standard teleportation via the the state $\varrho$. Here we adjust the
standard teleportation scheme \cite{Bennett_tel} so that it
provides perfect transmission for
the state of the form (\ref{max})
rather than for the true singlet one (as in original scheme).

The parameter $F$ for states will denote simply the fraction of the singlet
state given by $F(\varrho)=\langle \Psi_+|\varrho|\Psi_+\rangle$. For the
channels, we will denote by $F(\Lambda)$ the entanglement fidelity
of $\Lambda$ \cite{Schumacher} given exactly by $F(\varrho_\Lambda)$.
Then if one sent half of singlet down the channel, the entanglement fidelity
says how close is the output state to the input one. By definition, $F$ is
invariant under  the isomorphism (\ref{map}).

\section{Fidelity of teleportation and singlet fraction}
\label{fidelity-fraction}
In this section we  relate the optimal fidelity of
teleportation via a given mixed state to the maximal singlet fraction
attainable by means of {\it trace-preserving} LQCC operations.
This will lead
us in particular to the bound for fidelity of the conclusive teleportation
in the case of quasi-distillation. Our basic tool will be the novel twirling
technique \cite{xor} i.e. random application of $U\otimes U^*$
unitary transformations. We also introduce {\it twirling channels} which
is operation on channels analogous to twirling states.
\subsection{Teleportation}
Suppose that Alice and Bob share a pair of particles in a  given state
$\varrho$ acting on the Hilbert space
${\cal H}_A\otimes {\cal H}_B=C^d\otimes C^d$ and Alice has
a third particle in  unknown state $\psi\in {\cal H}_3=C^d$ to be teleported.
The standard teleportation scheme has been
described in the introduction.
The most general teleportation scheme
is that Bob and Alice given the particles in states
described above apply some trace-preserving
(hence without selection of ensemble) LQCC operation $\cal T$  to the
particles they share and the Alice's particle.
After the operation the state of Bob particle (from the pair) is to be
close to the unknown state of the third particle. The final state of Bob
particle is given by the following formula
\begin{equation}
\varrho^{\psi}_{Bob}={\rm Tr}_{3,A}\left[ {\cal T}(|\psi\rangle\langle \psi|\otimes
\varrho) \right].
\label{Bob}
\end{equation}
This establishes a quantum channel
$\Lambda_{{\cal T},\varrho}$
which maps the input state
(the state of the third particle) onto the output one -- the final state
of Bob particle
\begin{equation}
\Lambda_{{\cal T},\varrho}(|\psi\rangle\langle\psi|)=\varrho^{\psi}_{Bob}.
\end{equation}
This is a different  way of ascribing a channel to the given
state than the isomorphism (\ref{map}). It is determined by each established
teleportation protocol $\cal T$, and in contrast to the isomorphism,  it is
in general not a one-to-one association: two different states
$\varrho$ and $\varrho'$ can give the same
teleportation channel $\Lambda_{{\cal T},\varrho}=\Lambda_{{\cal T},\varrho'}$
(this was discussed in Ref. \cite{huge}).
The fidelity of a teleportation protocol $\cal T$ (via a given state
$\varrho$) is the fidelity of the arising channel
$f(\Lambda_{{\cal T}, \varrho})$.
According to our definition of $f(\varrho)$, given  in the previous section,
we have $f(\varrho)=f(\Lambda_{{\cal T}_0,\varrho})$, where ${\cal T}_0$ is
the standard teleportation protocol.
We must stress here  that, in general, we do not know whether
$\Lambda_{\varrho}=\Lambda_{{\cal T}, \varrho}$ for
some protocol $\cal T$, even if the protocol could {\it depend} on
$\varrho$. (thus we do not know whether the isomorphism implies
also the physical equivalence, see previous section). A particular
example where it is the case (depolarizing channel) will be discussed
subsequently (see subsection C).

Finally, let us consider the situation with restricted input i.e. if
$\dim {\cal H}_1=m<d$. Then we must work with some established embedding of
the space ${\cal H}_1$ into the Bob space ${\cal H}_2$ (of course
the very form of the embedding is here irrelevant) so that the formula
(\ref{fid2}) for fidelity is well defined.

\subsection{Noisy singlet}
Consider the one-parameter family of states \cite{xor}
given by
\begin{equation}
\varrho_p=pP_++(1-p){I\otimes I\over d^2},\quad 0\leq p \leq 1.
\label{ntryplet}
\end{equation}
We will call them noisy singlets. They are the most natural generalization
of the $2 \times 2$ Werner states \cite{Werner,Pop}.

Let us now calculate the two parameters $f$ and $F$. To calculate  $f$
consider the standard teleportation scheme
via the state. The scheme produces fidelity $1$ for singlet state.
For the completely random noise represented by the state
${I\otimes I\over d^2}$,
the average final state of Bob particle after the teleportation procedure
is equal to ${1\over d}I$ and does not depend on the unknown state to be
teleported. Then, in this case the fidelity amounts to $1\over d$.
Thus for the noisy singlet we obtain
\begin{equation}
f=p+(1-p){1\over d},\quad  {1\over d} \leq f \leq 1.
\label{f}
\end{equation}
The parameter $F$ amounts to
\begin{equation}
F=p+(1-p){1\over d^2},\quad {1\over d^2} \leq F \leq 1.
\label{F}
\end{equation}
The two parameters are related in the following way
\begin{equation}
f={Fd+1\over d+1}.
\label{fntryplet}
\end{equation}
We see that the noisy singlet is uniquely determined by any of those
parameters (so  one can use notation $\varrho_f$  or $\varrho_F$ if we use
one of these parametrizations).
The separability of the state $\varrho_{p}$ can be characterized in a
very clear way. Namely it is \cite{xor} separable if and only if
$0 \leq p \leq \frac{1}{d+1}$.
This is equivalent to ${1\over d^2} \leq F \leq \frac{1}{d}$
and ${1\over d} \leq f \leq \frac{2}{d+1}$.

Recall that the noisy singlets are the only states invariant
under $U\otimes U^*$
transformations \cite{xor}
(here the star denotes complex conjugation).
Any state $\varrho$, if subjected to
$U\otimes U^*$ twirling produces noisy singlet:
\begin{equation}
twirl(\varrho)\equiv
\int{\rm d}\, U U\otimes U^*\varrho U^{\dagger}\otimes U^{*\dagger}=
\varrho_F
\end{equation}
with $F={\rm Tr}(\varrho P_+)$. Thus the singlet fraction is
invariant under the twirling procedure.

\subsection{Depolarizing channel}
The depolarizing channel \cite{huge,xor} is defined as follows
\begin{equation}
\Lambda_{p}^{dep}(\sigma)=p \sigma+ (1-p)\frac{I}{d}
\label{dep}
\end{equation}
where $\sigma$ is the state acting on $C^{d}$.
From the formula (\ref{dep})
follows that with probability $p$ the channel does not affect the input
state, while
with probability $1-p$ it completely randomizes the input state.

Now if we  apply the considered channel to half of singlet we obtain the
noisy singlet with the same parameter $p$. Thus we have the equivalence
\begin{equation}
\Lambda^{dep}_p=\Lambda_{\varrho_p}
\end{equation}
Then it follows that $F(\Lambda^{dep}_p)=F(\varrho_p)$.
Even more,
we have {\it full} physical equivalence between depolarizing channel and
noisy singlet: the channel can be reproduced by standard teleportation applied
to noisy singlet (this is compatible with similar observation
in Ref.\cite{huge} for two-qubit case).
Thus we have the complete set of equivalences
\begin{equation}
\Lambda^{dep}_p=\Lambda_{\varrho_p}=\Lambda({\cal T}_{0}, \varrho_{p})
\end{equation}
To prove the last equality
consider the standard teleportation
scheme of an unknown state through the state
$\varrho_{p}$. As described in the previous subsection, with probability $p$
Bob will obtain the input state undisturbed, while with probability $1-p$
he will end up with totally mixed state $I/d$.

Thus any given input state $\sigma$
is in the process ${\cal T}_{0}$ transformed
into the state $p\sigma+(1-p)\frac{I}{d}=(I\otimes\Lambda_{p}^{dep})\sigma$.
Then it follows that also the parameter $f$ is the same for $\Lambda^{dep}_p$
and $\varrho_p$, so that the formulas (\ref{f}), (\ref{F}) hold also for
the depolarizing channel, and any of the parameters $F$ and $f$ determines it
uniquely.

\subsection{Twirling channels}
Here we will introduce an operation over the channels which is equivalent
to $U\otimes U^*$ twirling of states. Namely for any channel $\Lambda$ one can
consider a new one constructed in the following way. Given the incoming
particle, Alice subjects it to a random unitary transformation $U$, then sends
it through the channel and informs Bob, which unitary was applied.
Subsequently, Bob, received the particle, applies the inverse transformation
$U^\dagger$.

Now, we will show that, as expected,  the following lemma is true.

{\it Lemma 1.-}
Any channel $\Lambda$ subjected to the twirling procedure becomes
depolarizing channel with the same $F$ i.e. we have
\begin{equation}
twirl(\Lambda)=\Lambda^{dep}
\end{equation}
with $F(\Lambda^{dep})=F(\Lambda)$.

{\it Proof.-}
Let us first show that
\begin{equation}
\varrho_{twirl(\Lambda)}=twirl(\varrho_\Lambda)
\label{twirl-iso}
\end{equation}
which can be illustrated by means of the following commutative diagram
\begin{equation}
\begin{array}{ccccc}
      &\varrho     &\leftrightarrow & \Lambda_\varrho& \\
twirl & \downarrow &                &\downarrow      & twirl \\
      &\varrho_p   &\leftrightarrow & \Lambda_p^{dep}& \\
\end{array}
\end{equation}
where the arrows $\leftrightarrow$ denote the isomorphism (\ref{map}).
That the diagram commutes can be verified directly
\begin{eqnarray}
&\varrho_{twirl(\Lambda)}=(I\otimes twirl(\Lambda))P_+=&
\int{\rm d}\, U \left[ I\otimes U^\dagger I\otimes \Lambda(I\otimes U P_+
I\otimes U^\dagger)I\otimes U\right]=\nonumber\\
&&\int{\rm d}\, U \left[ I\otimes U^\dagger I\otimes \Lambda(U^T\otimes I P_+
U^*\otimes I)I\otimes U\right]=\nonumber\\
&&\int{\rm d}\, U \left[ U^{*\dagger}\otimes U^\dagger
(I\otimes \Lambda)P_+ U^*\otimes U\right]=\nonumber\\
&&\int{\rm d}\, U U^*\otimes U\varrho_{\Lambda} U^{*\dagger}\otimes U^{\dagger}=
twirl(\varrho_{\Lambda})
\end{eqnarray}
Here we used the identity $I\otimes AP_+=A^T\otimes I P_+$ \cite{Jozsa} and
the invariance of the Haar measure under Hermitian conjugation.

Now, applying the isomorphism (\ref{map}) we obtain that the channel
$twirl(\Lambda)$ is equal to the channel corresponding to the state
$twirl(\varrho_\Lambda)$. Since the latter is noisy singlet, then the channel
must be depolarizing one. Let us compute entanglement fidelity of
$twirl(\Lambda)$
\begin{equation}
F(twirl(\Lambda)\equiv F(\varrho_{twirl(\Lambda)})=F(twirl(\varrho_\Lambda))=
F(\varrho_\Lambda)\equiv F(\Lambda),
\end{equation}
where we used definition of $F$, its invariance under twirling states,
and the equality (\ref{twirl-iso}).
Hence the entanglement fidelity is  invariant under the twirling of channel.

We have also the following lemma.

{\it Lemma 2.-}  The channel fidelity $f$ is invariant under twirling.

\begin{equation}
f(\Lambda)=f(twirl(\Lambda)).
\end{equation}

{\it Proof.-} This follows from direct calculation of $f$.
Namely the formula (\ref{fid2})
can be rewritten as follows
\begin{equation}
f(\Lambda)=\int{\rm d}\ U {\rm Tr}\left(
U|\phi\rangle\langle\phi|U^{\dagger}
\Lambda(U|\phi\rangle\langle\phi|U^{\dagger})\right).
\label{fid3}
\end{equation}
where $\phi$ is an arbitrarily established
vector and the integral is performed over the uniform distribution on
the group $U(d)$ (proportional to the Haar measure).
Consequently, we have
\begin{eqnarray}
&&f(twirl(\Lambda))=\int{\rm d}\, U
{\rm Tr}\left(U|\phi\rangle\langle\phi|U^\dagger
\int {\rm d}\, V V^{\dagger}\Lambda(VU|\phi\rangle\langle
\phi|U^\dagger V^\dagger)V\right)=\nonumber\\
&&\int {\rm d} \, V \int {\rm d} \, U
{\rm Tr} \left(VU|\phi\rangle\langle\phi|U^\dagger V^\dagger
\Lambda(VU|\phi\rangle\langle\phi|U^{\dagger}V^{\dagger})\right)=\nonumber\\
&&\int {\rm d} \, V \int {\rm d} \, U
{\rm Tr}\left(U|\phi\rangle\langle\phi|U^\dagger
\Lambda(U|\phi\rangle\langle\phi|U^\dagger)\right)=\int {\rm d}\, V
f(\Lambda)=f(\Lambda)
\end{eqnarray}
with $V$ unitary and ${\rm d} V$ representing the integration over
Haar measure.

These two lemmas produce the following result.

{\it Proposition 1.-} For any channel $\Lambda$ one has
\begin{equation}
f(\Lambda)={F(\Lambda)d+1\over d+1}.
\end{equation}
{\it Proof.-}
The above equality is true for depolarizing channel, but as shown above both
$f$ and $F$ are invariants of twirling, so that it must be also true for
any channel.


\subsection{Teleportation and singlet fraction}

Here we will prove the main result of this section. Namely we will relate
the maximal fidelity of general teleportation scheme to the maximal possible
fraction of singlet attainable by means of trace-preserving LQCC operations.
This will reduce the problem of optimal teleportation scheme for a given mixed
state to the less complicated problem of increasing singlet fraction.

{\it Theorem.- }
Let $F_{max}$ be the maximal possible fidelity (overlap with the state	$P_+$)
which can be obtained from a given state $\varrho$ by means of
trace-preserving LQCC operation. Then the maximal  fidelity
$f_{max}$ of teleportation via the state $\varrho$ attainable
by means of trace-preserving LQCC operations
is equal to
\begin{equation}
f_{max}={F_{max}d+1\over d+1}.
\label{optimal}
\end{equation}

{\it Proof.-} First we will prove that $f_{max}\leq{F_{max}d+1\over d+1}$.
Suppose we have a teleportation channel of fidelity $f_{max}$.
From Prop. 1 it follows that entanglement fidelity $F$ of that channel
satisfies $f_{max}={Fd+1\over d+1}$. Then sending half of singlet down the
channel one produces state with $F$ satisfying relation (\ref{optimal})
and $F_{max}$ is at least equal to $F$.

Conversely, suppose that by trace-preserving LQCC operations a state $\varrho'$
of maximal $F$ has been obtained. Apply twirling to this state. The
resulting state is of the form (\ref{ntryplet}). Then the fidelity $f$ of
standard teleportation via this state satisfies the relation
(\ref{fntryplet}). Thus the standard teleportation will achieve the required
$f$ which ends the proof.

{\it Remark.-} It can be seen that we can assume that
the final singlet is less dimensional than the space
${\cal H}_1\otimes {\cal H}_2$.
For example, one can  consider maximal attainable fraction $F_m$ of
$m \times m $ singlet $|\Psi^{m}_+\rangle=\frac{1}{\sqrt{m}}
\sum_{i=1}^m|i\rangle|i\rangle$ where $m<d$.
In this case, the formula \ref{optimal} desribes the optimnal fidelity
of teleportation for restricted input (i.e. if the unknown state  comes
from the Hilbert space of dimension $m$ \cite{Lind_ost}.

\section{Optimal teleportation fidelity for bound entangled states}
\label{BE}

Here we will apply the results of the previous section to the question of
optimal fidelity of teleportation  via bound entangled (BE) states.
These are the ones which are entangled (are not mixture of product states)
but cannot be distilled \cite{bound}. Linden and Popescu asked the
question, whether the BE states
allow for better fidelity than the one of purely classical teleportation (i.e.
the one where Alice and Bob have no prior entanglement so that the quantum
information is sent via classical bits themselves). Positive answer to
the same question but in the context of states allowing local hidden variable
model \cite{Werner} allowed to obtain nonclassical features of the states
\cite{Pop}.  Now, for a class of BE states, those authors
obtained negative answer, taking into account more general teleportation scheme
than the original one (they allowed for arbitrary von Neumann measurement of
Alice).

Here, we are able to obtain the fully general answer. Namely we will show  that
{\it the optimal fidelity of teleportation via arbitrary BE state
is equal to the classical teleportation fidelity}.

Let us first derive the expression for the classical teleportation fidelity.
Due to the proved theorem,
it suffices to find maximal singlet fraction attainable
via classical communication. This, however, reduces to determining
the maximal possible singlet fraction of separable states. Of course,
it cannot be greater than $1/d$, since states with $F>1/d$ are entangled
(even free entangled i.e. distillable --- explicit distillation protocol
has been provided in Ref. \cite{xor}). On the other hand, as mentioned in the
previous section,  the noisy singlet with $F=1/d$ is separable.
Hence, applying the formula (\ref{optimal}) we obtain that the best
fidelity of teleportation via classical channel is given by $f_{cl}={2/(d+1})$.

Consider now the BE states. Since the states with $F>1/d$ are free entangled,
then the maximal possible $F$ for BE states is also $F=1/d$. Then
the maximal fidelity of teleportation via a given BE state is also
less than or equal to the $f_{cl}$. In fact it is equal, as having any BE
state one can simply get rid of it and perform classical teleportation attaining
the fidelity $f_{cl}$.

\section{Conclusive teleportation and increasing singlet fraction}
\label{conclusive}
Here we will consider the problem of conditional increasing of fidelity of
teleportation i.e. conclusive teleportation \cite{Tal}. By the
results of the section \ref{fidelity-fraction} this question will
be directly related to the problem of conditional increasing of singlet
fraction.

Suppose that Alice and Bob has a pair in state for which the optimal
teleportation fidelity is $f_0$. Suppose further, that the fidelity is too
poor for some Alice and Bob purposes. What they can  do to change the situation
is to perform the so-called conclusive teleportation. Namely, they
can perform some LQCC operation with two final outcomes 0 and 1. Obtained the
outcome 0 they fail and decide to discard the pair. If the outcome is 1 they
perform teleportation, and the fidelity is now much better that the initial
$f_0$. Of course, the price they must pay is that the probability of the
success (outcome 1) may be small. The scheme is illustrated on the figure 1.

A simple example is the following.
Suppose that Alice and Bob share a pair in pure state
$\psi=a|00\rangle+b|11\rangle$ which is nearly product
(i.e. $a$ is close to 1). Then the standard teleportation scheme  provides
a rather  poor fidelity $f={2\over 3} {a^3-b^3\over a-b}$
\cite{Gisin_tel,tel}. However, Alice can subject her particle to filtering
procedure \cite{Gisin,conc} described by the operation
\begin{equation}
\Lambda=W(\cdot)W^\dagger +V(\cdot)V^\dagger
\end{equation}
with $W={\rm diag}(b,a)$, $V={\rm diag}(a,b)$. Here the outcome 1 (success)
correspond to operator $W$. Indeed, if this outcome was obtained, the state
collapses to the singlet one
\begin{equation}
\tilde\psi={W\otimes I\psi\over ||W\otimes
I\psi||}={1\over\sqrt2}(|00\rangle+|11\rangle)
\end{equation}
Then, in this case  perfect teleportation can be performed. Thus, if Alice and
Bob teleported directly via the initial state, they would obtain
a very poor performance. Now, they have a small, but nonzero chance of
performing perfect teleportation.

The main questions concerning the above scheme of conclusive teleportation
are the  following.  Which states can provide perfect
conclusive teleportation?
More precisely, given a state $\varrho$, does there exist a nonzero
probability  $p$ of success, for which Alice and Bob end up with pure singlet?
Confining now to the class of states which
cannot be converted into pure singlets
one could ask: how large fidelity can be obtained? As we will see,
{\it the fact that perfect conclusive teleportation is impossible does not,  in
general, mean that there is some fidelity threshold $C<1$ which cannot
be exceeded}.

To analyse the above questions, we will apply the tools worked out in previous
sections. Namely, there we have reduced the problem of optimal fidelity of
teleportation  to the problem of optimal increasing of singlet fraction.
Let us now apply this result to the present situation. Namely
for a given probability $p$ let $f_p$ denote the maximal fidelity of
conclusive teleportation with this probability of success. From the theorem
it follows that $f_p={F_pd+1\over d+1}$  where $F_d$ is the maximal singlet
fraction attainable with probability $p$ of success. So to obtain results
concerning fidelity of teleportation we do not need to consider conclusive
teleportation scheme but the much simpler scheme of conclusive increasing
singlet fraction. The scheme is illustrated on the figure 2.

Again Alice and Bob, perform some LQCC operation with two outcomes.  The
outcome 0 denotes failure, while obtained the other outcome, Alice and Bob
have the final state of higher fraction of singlet than the initial one.
Now the question concerning the fidelity of teleportation can be reformulated in
the following way. For which states perfect singlet can be obtained?
For which states arbitrarily high fraction of singlet can be obtained?
Finally, for which states there is a threshold for singlet fraction which
cannot be exceeded?

Applying now the known results concerning increasing singlet fraction
\cite{Kent,Linden} we obtain that there is such a treshold for the states
of full rank (i.e. with eigenvalues non-vanishing).
However,
we will provide the class of states of low rank, which do not allow for
perfect conclusive teleportation, but still {\it arbitrarily high fidelity}
can be obtained with nonzero probability (the latter depend on how high
fidelity we would like to have). We will also provide a class of states of
low rank for which we prove that the threshold exists. As we will see the
proof is surprisingly complicated. Then the problem
of determining whether a given state has an ultimate threshold
for conclusive teleportation becomes highly nontrivial.

The above problems are closely related to the problem of distillation
\cite{Bennett_pur}
by means of non-collective operations \cite{Linden}. Namely, if for some
state it
is possible to obtain conclusively perfect singlets, then we have in
fact a protocol of distillation, because we obtain a nonzero rate of
produced singlets. Then such a state is non-collectively distillable.
In the case, where pure singlets cannot be produced,
the non-collective operations cannot produce nonzero asymptotic yield.
If still an arbitrary high singlet fraction can be obtained, we will
call the state non-collectively {\it quasi-distillable} (as in this paper we
deal only with non-collective protocols, so we will say briefly
quasi-distillable). The states which have the ultimate threshold of fraction of
singlet we call non-quasi-distillable. In subsequent sections we will define
the notions more precisely, and we will consider the relevant examples.


\section{Non-collective $ m \times m $ distillation}
\label{distillation}
Let $P_+^{m}= |\Psi_{+}^{m} \rangle \langle \Psi_{+}^{m}|$.
As mentioned in introduction, following the ideas
presented in the papers \cite{Linden}, \cite{Kent}
we use the following definition of noncollective distillation.

{\it Definition .-}
One says that the $N \times M$ state $\varrho$
can be $m \times m$ non-collectively distilled
iff there exist operators $A$, $B$ such that
\begin{equation}
\frac{A\otimes B \varrho A^{\dagger} \otimes B^{\dagger}
}{Tr(A\otimes B \varrho A^{\dagger} \otimes B^{\dagger})}=P_{+}^{m}
\label{ND}
\end{equation}

We shall need other notions yet.

{\it Definitions .-}
(i) If the state has the Schmidt decomposition
\begin{equation}
\Psi=\sum_{i=0}^{m-1}a_{i} | f_{i}'\rangle | f_{i}'' \rangle, \quad
a_{i}\neq 0
\label{dec}
\end{equation}
then we shall call the number $m$ {\it the Schmidt rank of state $\Psi$}
and denote it by $r_s(\Psi)$.

(ii) We also shall call {\it product $n \times m$ projection}
the product projection
$ P \otimes Q $
where the ranks of the projections $P$, $Q$ are
$n$, $m$ respectively.
Hilbert subspace  of the space ${\cal H}$
corresponding to any such projection
we shall call {\it product $n \times m$  subspace}.

Here we simply characterize the states which
can be non-collectively distilled.

{\it Proposition 2.-} A given $N \times M$ state $\varrho$
is $m \times m$ non-collectively distillable
iff there exists $m \times m$ product projection
$P \otimes Q$ such that $P \otimes Q \varrho P \otimes Q$ is
some pure (possibly unnormalized) projector of Schmidt rank $m$.

{\it Proof .- }

Consider the given state $\varrho$
that is non-collectively distillable.
It means that there exists some $A \otimes B$
such that
\begin{equation}
A \otimes B \varrho A^{\dagger} \otimes B^{\dagger}=|\phi \rangle \langle \phi|
\label{czysty}
\end{equation}
and $|\phi \rangle$ is (possibly unnormalized) maximally
entangled vector of rank $m$.
Note, that one can restrict to
hermitian $A$, $B$. It
follows from two simple facts:
(i) for any $A$, $B$ there exist $\tilde{A}$,
$\tilde{B}$, and unitary $U_{A}$, $U_{B}$ such that  $A \otimes B=
\tilde{A} U_A\otimes \tilde{B} U_{B}$,
(ii) product unitary transformation
$U_{A} \otimes U_{B}$ does not change
Schmidt rank.

Consider now hermitian $A$, $B$ satisfying (\ref{czysty}).
One can invert them on their supports:
$A^{-1}A=P_{A}$, $B^{-1}B=P_{B}$ where $P_{A}$, $P_{B}$
are projections onto the supports of $A$, $B$. Consider a new vector given by
$|\psi\rangle=A^{-1}\otimes B^{-1} |\phi\rangle$. As no product operator
can increase the Schmidt rank \cite{Hoi-Kwong} we have
$r_s(\psi)\leq r_s(\phi)$. Since $|\phi\rangle=A\otimes B |\psi\rangle$, we
obtain that, in fact, $r_s(\psi)=r_s(\phi)=m$. Also, by definition of $P_A$,
$P_B$, $A^{-1}$ and $B^{-1}$ one has
\begin{equation}
|\psi\rangle\langle\psi|=P_A\otimes P_B \varrho P_A\otimes P_B
\end{equation}
The projector $P_A\otimes P_B$ must have at least rank  $m\times m$ (otherwise
$r_s(\psi)$ would have to be less than $m$). If it has grater rank, then
one can easily find (via Schmidt decomposition of $\psi$) the new projectors
$P_A'\otimes P_B'$ of rank $m\times m$ that still convert $\varrho$ into
$|\psi\rangle\langle\psi|$. Thus, if $P_A\otimes P_B$ has rank $m \times m$
then we take $P=P_A$, $Q=P_B$, otherwise $P=P_A'$, $Q=P_B'$.

Suppose now, conversely, that there exists $P \otimes Q$ of rank $m\times m$
such that $P\otimes Q\varrho P\otimes Q=|\psi\rangle\langle\psi|$ with
$r_s(\psi)=m$. Then $\psi$ is of the form (\ref{dec}). Now, taking
$A=P$, $B=VQ$, with $\langle f_i''|V|f_j''\rangle=(1/a_i) \delta_{ij}$ (see
(\ref{dec})), one
obtains that $A\otimes B\varrho A^{\dagger}\otimes B^{\dagger}$ is maximally
entangled state (of Schmidt rank $m$, of course).
Note that the operator $V$ plays the role of the
suitable local filter \cite{conc,Gisin}.
Now, applying suitable product uunitary transormation, we
obtain (after normalization) the
desired state $P_+^{m}$. This ends the proof.

%
%

Note that the above proposition provides the necessary and sufficient
condition for noncollective distillation, which obviously
does not automatically provide the best way to distill the state.
From the proposition it follows directly that {\it no
mixed state of $d \times d$ system can be converted
into the maximally entangled state of the system}.


This is possible however for many states of the system
$N \times M$ , $M > N$.
Simple examples of such states are
the states of the form $p|\Psi_{+} \rangle \langle \Psi_{+}|
+ (1-p) \varrho'$ with
reduced density matrix $\varrho'_{2}$ of the matrix $\varrho'$
orthogonal to the projector $P=\sum_{i=0}^{N-1}|i\rangle \langle i|$.
The corresponding operators $A$, $B$ turning  such states
into maximally entangled state are $A=I$ (identity operator on the
first subsystem) and $B=P$.

\section{Non-collective quasi-distillation}
\label{quasi-distillation}
One can ask whether it is possible by means of LQCC operations
to make $F_{m}$ arbitrary close to 1 with nonzero probability
even if the non-collective distillation
(\ref{ND}) is impossible.
In fact one can imagine a sequence of LQCC operations
producing better and better $F$ but with the probability tending to zero.
Then the corresponding  denominators of the expression (\ref{ND})
converge to zero, so that  the hypothetical limiting operation
does not exist.
It corresponds to the existence of $A_{n}$, $B_{n}$ such that
\begin{equation}
\frac{A_{n}\otimes B_{n} \varrho A_{n}^{\dagger} \otimes B_{n}^{\dagger}
}{Tr(A_{n}\otimes B_{n} \varrho A_{n}^{\dagger} \otimes B_{n}^{\dagger})}
\mathop{\longrightarrow}\limits_{}^{n \rightarrow \infty}
P_{+}^{m}
\label{NQD}
\end{equation}
The existence of such the operators we shall call
the noncollective {\it quasi-distillation}
as we allow corresponding
sequence of probabilities $p_{n}=
Tr(A_{n}\otimes B_{n} \varrho A_{n}^{\dagger} \otimes B_{n}^{\dagger})$
to decrease to zero. It means that if such noncollective
operations were performed on many pairs of particles then,
unlike in the original distillation scheme \cite{Bennett_pur}, one would obtain
zero yield \cite{Bennett_pur,huge} of pure singlet states $P_{+}^{m}$.

Now we are in position to present an example of the quasi-distillation
process. In this section we shall focus on the quasi-distillation
of the $d \times d$ system
to the maximally entangled state $P_+=P_+^d$ (not to $P_+^m$ with $m<d$).
To be specific, we will deal with the case $d=3$.

The mixed state which can exhibit arbitrary high fidelity $F$ after
non-collective local filtering is the following:
\begin{equation}
\sigma_F=FP_{+}+(1-F)|01\rangle \langle 01|, \ \ 0<F<1.
\label{st1}
\end{equation}
Following our remarks from previous section
we know that the state, as a mixed one,  cannot be distilled
to the maximally entangled state $P_{+}$ .
According to the formula (\ref{optimal})
this means that it is impossible to teleport with
the fidelity $f=1$ via the state $\sigma_F$.

However it can be easily seen that the operations:
\begin{eqnarray}
A_{n}\equiv diag [\frac{1}{n},1,1], \ \
B_{n}\equiv diag [1,\frac{1}{n},\frac{1}{n}]
\end{eqnarray}
allow for quasi-distillation process (\ref{NQD}).
Indeed, then
\begin{equation}
A_{n}\otimes B_{n} \sigma A^{\dagger}_{n} \otimes
B^{\dagger}_{n}=\frac{1}{n}[ F P_{+} +\frac{1-F}{n}|01\rangle \langle 01| ]
\end{equation}
which after suitable normalisation leads to the desired result.
The key point is that in the letter example
the normalizing factor of the new state
converges to zero i.e. we have increasing of the fidelity
of the output state but, at the same time,
the probability of obtaining  this output state
decreases to zero.
Now, according to (\ref{optimal})
it means that it is possible to teleport with
the fidelity $f$ arbitarily close to
unity, although the value  $f=1$ can not be achieved.

Now we shall see how the above situation dramatically
change under seemingly not strong modifications of the input state (\ref{st1}).
For this purpose consider the following state \cite{aktywacja}
\begin{equation}
\varrho_{F}=FP_{+}+\frac{(1-F)}{3}(|01\rangle \langle 01|  + |12\rangle \langle 12|
+ |20\rangle \langle 20|), \ \ 0<F<1.
\label{st2}
\end{equation}
For  convenience let us introduce the notation $\Theta_{n}(\sigma)=
A_{n}\otimes B_{n} \sigma A_{n}^{\dagger} \otimes B_{n}^{\dagger}$,
$\Theta(\sigma) = A\otimes B \sigma A^{\dagger} \otimes B^{\dagger}$,
and  $\langle \omega \rangle = Tr(\omega)$.
The same arguments as before lead to the conclusion
that there are  no operators $A$, $B$ such that
$\frac{\Theta(\varrho_{F})}{\langle \Theta(\varrho_{F}) \rangle}=P_{+}$.
We will show that for the considered  state, unlike for $\sigma_F$,
even the second, weaker form of distillation of entanglement is impossible.
Let us assume, on the contrary, that (\ref{NQD}) is possible.
Then the output states can be written as  convex combinations of two states,
the second one of them being certainly {\it separable}
\begin{eqnarray}
\frac{\Theta_{n}( \varrho_{F})
}{\langle \Theta_{n}( \varrho_{F})\rangle}=
\left[F \frac{\langle\Theta_{n}(P_{+})\rangle
}{\langle\Theta_{n}(\varrho_{F})\rangle}\right]
\frac{\Theta_{n}(P_{+})}{
\langle\Theta_{n}(P_{+})\rangle} \nonumber \\
+ \left[(1-F)\frac{\langle\Theta_{n}(\sigma_{+})\rangle
}{\langle\Theta_{n}(\varrho_{F})\rangle}\right]
\frac{\Theta_{n}
(\sigma_{+})}{\langle\Theta_{n}(\sigma_{+})\rangle}
\label{limit}
\end{eqnarray}
where $\sigma_+={1\over3}(|01\rangle\langle01|+|12\rangle\langle
12|+|20\rangle\langle 20|)$; the weights at both states have been put
into squared brackets
(assume, for a while, that for all $n$, we have $\Theta_n(P_+)\not=0$ and
$\Theta_n(\sigma_+)\not=0$).
Since the limit state must be pure entangled state, the second weight
must converge to zero. Otherwise some of its subsequences
would converge to the weight $w_{2}>0$
(recall that any bounded sequence has a convergent subsequence).
As both the set of states and the set of separable states are compact it
would lead to the conclusion that the limit state of the sequence
(\ref{limit}) includes some separable state with the nonzero
weight $w_{2}$. Such a  state certainly could not
be the pure entangled one.
Thus the weight
$(1-F)\frac{\langle\Theta_{n}(\sigma_{+})\rangle
}{\langle\Theta_{n}(\varrho_{F})\rangle} $
must converge to zero.
Together with the normalization condition it implies
immediately that the weight at the state
$\frac{\Theta_{n}(P_{+})}{
\langle\Theta_{n}(P_{+})\rangle}$
must converge to unity.
Hence we have
\begin{equation}
\frac{\Theta_{n}(P_{+})}{\langle\Theta_{n}(P_{+})\rangle}
\mathop{\longrightarrow}\limits_{}^{n \rightarrow \infty} P_{+}.
\label{cond2}
\end{equation}
We also obtain  that  the ratio of the second weight to the first one
must vanish in the limit of large $n$. This leads to
the condition
\begin{equation}
\frac{\langle \Theta_{n}(\sigma_{+})\rangle}{
\langle\Theta_{n}(P_{+})\rangle}
\mathop{\longrightarrow}\limits_{}^{n \rightarrow \infty} 0
\label{cond1}
\end{equation}


Subsequently, we shall show that satisfaction of the condition
(\ref{cond2}) is impossible if only (\ref{cond1})
is satisfied.
Let us introduce the notation
$|a_{i}^{n} \rangle= A_{n}|i \rangle / \sqrt[4]{\langle\Theta_{n}(P_{+})\rangle} $,
$|b_{i}^{n} \rangle= B_{n}|i \rangle / \sqrt[4]{\langle\Theta_{n}(P_{+})\rangle} $,
$i=1, 2, 3$.
Then the requirement (\ref{cond2}) can be rewritten as
\begin{eqnarray}
\Psi_{n} =
\frac{1}{\sqrt{3}}(|a_{0}^{n}b_{0}^{n} \rangle
+|a_{1}^{n}b_{1}^{n} \rangle  +
|a_{2}^{n}b_{2}^{n} \rangle)
\mathop{\longrightarrow}\limits_{}^{n \rightarrow \infty}
 \frac{1}{\sqrt{3}}(|00 \rangle
+|11\rangle  + |22\rangle ).
\label{cond2'}
\end{eqnarray}
But, at the same time, calculating that
$\langle\Theta_{n}(\sigma_{+}) \rangle=Tr(
\Theta_{n}(\sigma_{+}) )=
{1\over 3}\sum_{i=0}^{2} || A_{n}|i \rangle ||^{2}
||B_{n}| i \oplus 1 \rangle ||^{2}$ (here $x\oplus y= (x+y)\,{\rm mod}\, 3$)
we obtain via (\ref{cond1}) that
$\lim_{n \rightarrow \infty} \sum_{i=0}^{2} || |a_{i}^{n} \rangle ||^{2}
||b_{i \oplus 1}^{n} \rangle ||^{2}=0$.
The latter is the sum of three nonnegative sequences, so that
any of them must converge to zero independently.
Taking their square roots and multiplying them by each other we obtain,
after suitable reordering, that
\begin{equation}
\lim_{n \rightarrow \infty} (||a_{0}^{n}|| ||b_{0}^{n}||)
(||a_{1}^{n}|| ||b_{1}^{n}||)
(||a_{2}^{n}|| ||b_{2}^{n}||)=0
\end{equation}

Vanishing of this limit, which is a product of three positive sequences
implies that at least one of
them, say $||a_{0}^{n}|| ||b_{0}^{n}||$, must
converge to zero.
But it means, turning back to (\ref{cond2'}),
that $\lim_{n \rightarrow \infty } \Psi_{n} =\lim_{n\rightarrow\infty}
\frac{1}{\sqrt{3}}(
|a_{1}^{n}b_{1}^{n} \rangle  + |a_{2}^{n}b_{2}^{n} \rangle)$.
This limit vector obviously cannot be singlet state (as
(\ref{limit}) requires) because its
Schmidt decomposition can consist of at most two terms
(it can be easily seen by looking at
the spectrum of the corresponding reduced density matrix).
In this way we have obtained the required contradiction.

Finally, note that if the condition of nonvanishing of $\Theta_n(P_+)$
and $\Theta_n(\sigma_+)$ for all $n$ is not satisfied, the result is still
valid. Indeed, if for all but finite number of components of the sequence
$\Theta_n(P_+)$ vanish, then the limit state is separable (hence certainly
cannot be quasi distilled). If the same holds for the sequence
$\Theta_n(\sigma_+)$, then the state (\ref{limit}) consits
only of the first term and the proof still applies (the limits containing
$\Theta_n(\sigma_+)$ can be replaced by zeros). If none of the above
conditions is fulfilled, one can take a subsequence
$\Theta_{n_k}(P_+)$ and $\Theta_{n_k}(\sigma_+)$ with all components
nonvanishing and apply the proof to the subsequence.

Thus we have proved that for the considered state
the process (\ref{NQD}) is impossible.
In other words no LQCC operations
performed on the (\ref{st2}) state  can increase the fidelity
$F(\varrho_{F})$ upon some $C<1$.
Following the results of section \ref{fidelity-fraction} it means that
the conclusive
teleportation of the spin-1 state through the state $\varrho_{F}$ can
produce the fidelity of transmission at most equal
to $f_{max}=\frac{Cd+1}{d+1}$.

\section{Summary and conclusion}
We have developed the correspondence between states and channels.
In particular we have exploited the equivalence between
$U \otimes U^{*}$ invariant states and the  generalized depolarizing
channel to provide  a relation between the optimal fidelity of teleportation
and the maximal attainable singlet  fraction.
If the maximal fraction $F$ of singlet 
obtained from the initial two spin-s
state $\varrho$ by means of trace-preserving LQCC operations is
equal to $F_{max}$ then the best possible transmission fidelity
$f$ of teleportation via state $\varrho$
is $f_{max}=\frac{F_{max}d+1}{d+1}$. This result was applied to the case of
conclusive teleportation.
It gives the answer to the question announced at the beginning of
this paper. Namely, if Alice wants to teleport
only if she knows that the transfer fidelity is
better than $f_{0}$ then the state {\it must} admit
an LQCC operation converting  it (possibly with some
probability) into the state with singlet fraction greater than
$F_{0} =(1+\frac{1}{d})f_{0}-\frac{1}{d}$.

It is interesting that the result does not depend on
the kind of teleportation scheme and at the same
time it involves an quantity $F$ which
measures the degree of  overlap of the channel state with
the singlet  state.
The quantity was originally associated with the standard
teleportation scheme \cite{tel}.
Moreover, in this scheme Alice performs her complete
measurement (required as a step of the scheme)
in maximally entangled basis. It suggests that the standard
teleportation scheme might be optimal.

As  the concept of conclusive teleportation
with good fidelity appeared to be connected  with
the possibility of increasing of the fidelity $F$,
we have considered the problem
of noncollective distillation of the mixed state of two
component system. It involves a conversion
(by means of noncollective LQCC operations) of some
$d \times n$ ($d\geq n $) channel state
into the maximally entangled state $P_{+}$ or  its
$m \times m$ counterparts ($m<d$).
We have shown that
the first kind of conversion is suppressed  for
mixed states (this generalizes the result for $2\times 2$
singlets \cite{Kent}). Thus there is an important difference
between  $d \times d$
mixed and pure states as there are some pure states
which can be converted in such a way (see \cite{xor}).
The states for which the second kind of
conversion is possible have been characterized and
the possibility of conversion of some $d \times n $ ($d<n$) states
into the singlet state has been pointed out.

Then we have introduced the concept of quasi-distillation
which means,by definition, the possibility of making via LQCC process
the quantity $F$ arbitrary close
to unity with nonzero probability but with the latter allowed to
depend on the desired $F$ of output state.
We focused on the {\it noncollective} quasi-distillation which is
the possibility of making the fidelity $F$ of the
state arbitrary
close to unity in {\it noncollective LQCC
process}.
It turned out that sometimes despite that a state
cannot be non-collectively distilled it allows for
noncollective quasi-distillation. The key point is that the probability
of achieving of the required fidelity is the less the higher the fidelity is.
The examples of states which are  quasi-distillable but not
distillable via noncollective processes have been provided.
They show that impossiblity of perfect tleleportation
sometimes does  not implies the
treshold for fidelity of the conclusive teleportation
i.e. sometimes it is still possible to teleport with
$f$ arbitrarily close to unity.

Subsequently some modifications of those states
have been considered that do not fall into the classes considered
so far in Refs.\cite{Linden,Kent} and
for which the approach proposed in Ref. \cite{Kent} cannot be
applied due to low rank of the matrix of the state.
Nevertheless we have shown by different technique
that those states are not even quasi-distillable \cite{aktywacja}.
One of the main results of this paper is
the conclusion that, in the noncollective LQCC
operations regime, any state which cannot be quasi-distilled
{\it never} allow for conclusive teleportation
with the fidelity better than some boundary value $f_{max}$.
This result has been achieved by means of general
approach including {\it all} possible teleportation schemes.
It was possible due to the application
of the isomorphism between states and channels which
seems to be promising technique in quantum information theory.

We thank Noah Linden and Sandu Popescu for sending us their
manuscript on bound entanglement and teleportation \cite{Lind_ost} and
for useful comments. The work is supported by
Polish Committee for Scientific Research, Contract No. 2 P03B 103 16.
M. H. and P. H. kindly acknowledge the support
from the Foundation for Polish Science.

\begin{figure}
\caption[Conclusive teleportation]{Conclusive teleportation. Starting
with a weakly entangled pair Alice and Bob prepare with probability p
a strongly entangled pair and then perform teleportation.}
\caption[Conclusive increasing singlet fraction]
{Conclusive increasing singlet fraction. Alice and Bob with probability p of
success obtain a state with higher singlet fraction than the one of the
initial state.}
\end{figure}


\begin{references}
\bibitem{CP}
Recall that the map $\Lambda$ acting on
spin-$s$ states is called completely positive,
if the new map $\Lambda \otimes I$ acting
on the states of the system of two particles, one
of spin $s$ and the other of spin $s'$,
is positive for any second system
spin $s'$. We also call $\Lambda$ trace-preserving
if it preserves trace of all mapped matrices.
\end{references}
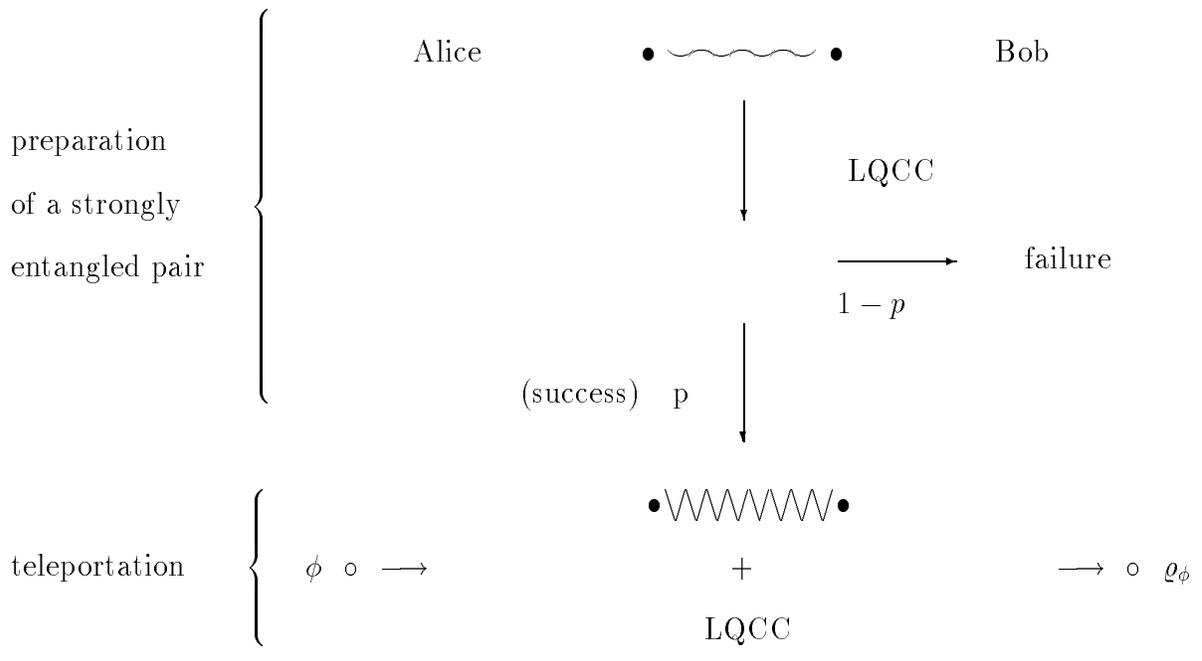
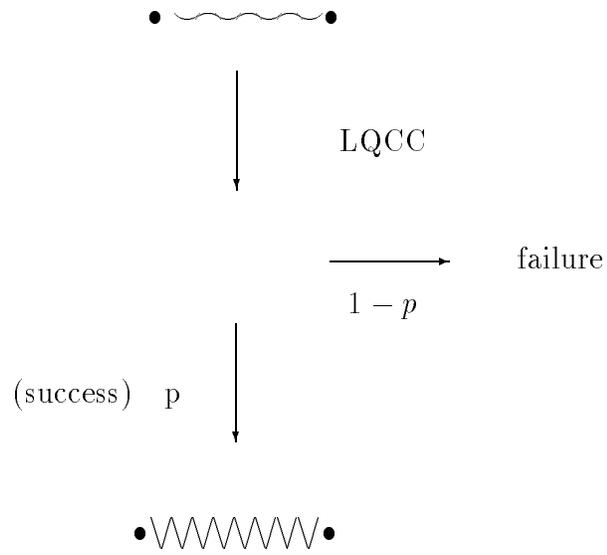
\end{document}